\newcommand{\bea}{\begin{eqnarray}}
\newcommand{\eea}{\end{eqnarray}}
\documentclass[aps,pra,preprint,amsmath,amssymb]{revtex4}
\usepackage{epsfig}
\usepackage{times}
\usepackage{mathrsfs}
\usepackage{graphicx}
\usepackage{dcolumn}
\usepackage{color}
\usepackage{bm}
\usepackage{epstopdf}
\usepackage{appendix}
\usepackage{amsmath,amssymb}
\begin{document}
\title{Finite-range model potentials for resonant interactions}
\author{Bimalendu Deb$^{1,2}$}
\email[]{}
\affiliation{}
\author{}
\affiliation{}
\author{}
\affiliation{$^1$ Department of Materials Science, $^2$ Raman Centre for Atomic, Molecular and Optical Sciences,
Indian Association for the Cultivation of Science,
 Jadavpur, Kolkata 700032. INDIA.}

\begin{abstract}
We show that it is possible to model two-body resonant interactions at low energy with a class of  finite-range potentials 
based on the methods of Jost and Kohn. These potentials are expressed in terms of the effective range $r_0$
and the $s$-wave scattering length $a_s$. We derive continuum solutions of these potentials.   
By writing 
$V_{\pm}(r) = V_{0}(r) + V_{\pm}^{\epsilon}(r)$,  where the sign +(-) refers to positive(negative) scattering length, $ V_{0}(r)$ 
is of the form of P\"{o}schl-Teller  potential and  $V_{\pm}^{\epsilon}$ is expressed as a power series of the small 
parameter $\epsilon = (\sqrt{1 - 2 r_0/a_s})^{-1} - 1 $ when  $a_s$ is large, we derive 
Green function of $V_{0}(r)$. Using the Green function, 
 solutions of $V_{\pm}(r)$ for $|a_s| >\!> r_0$ can be obtained numerically by treating $V_{\pm}^{\epsilon}(r)$ 
 as a perturbation.  
We describe the threshold behavior of 
scattering phase shift for  $V_{0}(r)$. This study may be important for developing a  better understanding of physics 
of strongly interacting ultracold atomic gases with tunable interactions.  
\end{abstract}

\maketitle


\section{Introduction} 

Physics of interacting many-particle systems is described, to a first approximation, in terms of a mean-field 
potential.  At low energy, there is a well-known mean-field potential of zero range. This is  known as  
Fermi's contact potential or pseudopotential  \cite{fermi:pseodopot},  usually expressed as a delta function with 
the potential strength being proportional to 
$s$-wave scattering length $a_s$ only.     
For such a delta-type potential, 
 $s$-wave scattering wave function at zero energy becomes singular at $r=0$, where $r$ is 
 the separation between two particles.  To circumvent this problem, contact interaction is replaced  
with a suitable  boundary 
condition at $r = 0$ in the Schr\"{o}dinger equation resulting in the regularized pseudopotential 
\cite{fermi2,breit1,book:blatt,pr:1957:huang-yang} 
\bea 
V({\mathbf r}) = \frac{4 \pi \hbar^2 a_s}{2\mu} \delta (\mathbf{r}) \frac{\partial}{\partial r} r 
\label{pseudo}
\eea 
where $\mu$ is the reduced mass of the two particles. The  zero-range potential 
method  is applicable to low temperature dilute atomic or 
molecular gases whose two-body interactions  obey Wigner's  
threshold laws \cite{book:mott-massey}. For two spherically symmetric atoms,  
Wigner's  threshold laws  dictate that  $a_s$ 
becomes an energy-independent parameter. The $s$-wave scattering amplitude then 
reduces to the form 
$ 
f_0(k) \simeq  - \frac{a_s}{1 + i k a_s} $
which is  $f_0(k) \simeq -a_s$ for $k |a_s| <\! < 1$.  
However, when a quasibound or bound state exists near the threshold of a potential or a scattering resonance occurs 
at low energy, $a_s$ becomes large 
leading  to the breakdown of the condition $k |a_s| <\!< 1$. 
Wigner's threshold laws do not apply when  a resonance occurs near zero energy.
As a consequence, 
Fermi's pseudopotential approach  becomes inadequate to describe collision or many-body physics near resonances.

The 
main purpose of this paper is to seek a finite-range model potential that can describe  resonant interactions.
Resonances and physical phenomena related to resonant interactions  are ubiquitous in physics and chemistry. 
In ultracold atoms, Feshbach resonances can be  induced by external magnetic \cite{rmp:2006:kohler,rmp:2010:chin} 
or optical \cite{prl:1996:fedichev,prl:2009:deb,rmp:2006:pa} or both magnetic and optical fields \cite{jpb:2010:deb}.
 In current literature, many-body phenomena in ultracold atomic gases near a Feshbach 
resonance are by and large described in terms of contact potential \cite{rmp:2008:bloch}. 
However, the effective range and energy dependence of many-body 
phenomena can hardly be ignored when a resonance occurs.
It is theoretically shown that  for $a_s >\!> R_{{\rm vdW}}$
when $a_s$ is much larger than the characteristic length scale 
$R_{{\rm vdW}} = (2 \mu C_6/\hbar^2)^{1/4}/2$ of van der Waals potential, the effective range $r_0$ is about 
three times $R_{{\rm vdW}}$ \cite{gao:1998,flambaum:1999}. The values of $R_{{\rm vdW}}$ for different alkali atoms  
used in  cold atom experiments are tabulated in Ref. \cite{rmp:2010:chin}. For $a_s >\!> R_{{\rm vdW}}$, the effective range 
$r_0$ of  
a potential with van der Waals dispersion at long separation will be several tens of Bohr radius ($a_0$).
This means $r_0$ is  large compared to typical molecular equilibrium positions. 
It has been recently shown that the  range and energy dependence 
of magnetic Feshbach resonances of ultracold atoms are quite important \cite{prl:ohara:2012,pra:2014:hutson}. Particularly in case of narrow Feshbach resonances, 
$r_0$ can be very large, even of the order of hundred or thousand $a_0$. 
Currently, attempts are being  made to construct  an improved model potential 
by resorting to a toy model \cite{pra:2014:ketterle} or 
contact potentials with  energy-dependent scattering length \cite{pra:2002:bolda}.  
The fact that a contact interaction  can not accurately describe resonant interactions
calls for the formulation of a better model potential that can take into account both the finite range  and the energy dependence of 
scattering amplitudes. 

Here we show that one can deduce, based on the methods of Jost and Kohn \cite{PHYSREV:Kohn:1952,kohn:1953}, 
a class of finite-range model potentials that can 
account accurately for $s$-wave resonances at low energy.  
We demonstrate that, for large $a_s$, the model potentials reduce to P\"{o}schl-Teller form \cite{ptpot} that 
can admit  
 analytical solutions. Using these analytical solutions, we 
construct Green function that can be used for solving the full potential numerically. 
 The primary physical motivations behind this work  is to establish the connections of strongly or unitrity-limted 
interactions with the exactly solvable P\"{o}schl-Teller potentials. 
This study showing that the unitarity-limited interactions can be treated with exact analytical 
solutions may be important for 
gaining new insight into the  physics of strongly interacting ultracold atomic gases. There is another motivation behind undertaking 
this work. In a seminal paper  Butsch {\it et al.}  \cite{foundphys:1998:butsch} have presented exact solutions  
of two identical cold atoms interacting via regularized contact potential in  one, two 
and three dimensional harmonic traps. 
However, as expected,  these exact solutions will not apply to  strongly interacting atoms in traps due to finite-range effects. 
It is therefore 
worth seeking exact solutions 
of two atoms interacting via the finite-range model potentials with large scattering length in a harmonic trap. Such solutions will 
lead to new insight into physics of strongly interacting systems in confined or low-dimensional space. In this paper we do not make 
any attempt to find solutions of two trapped atoms interacting with finite-range potentials. However, the finite-range potentials 
and their exact solutions presented in this paper may serve as a precursor towards generalizing the results of  
 Butsch {\it et al.}  \cite{foundphys:1998:butsch} for finite-range potentials. 

This paper is organized in the following way. In the next section, we present our proposed finite-range potentials for resonant interactions and discuss 
how they are obtained. In Sec. 3, we present continuum and bound state solutions of the potentials. Analytical and numerical results are discussed in Sec. 4. The 
paper is concluded in Sec. 5. 

\section{Finite-range model potentials}
The model potentials we consider are some variants of Bargmann potentials \cite{book:newton} and derived 
using effective range expansion of phase shift. The connections between phase shifts and potentials were 
first rigorously established by Bargmann \cite{bargmann:rmp:1949}. 
The early works on the method derivations of finite-range potential from the phase shift 
data were carried out by   Gel'fand and Levitan \cite{Levitan:1951}, 
Jost and Kohn \cite{PHYSREV:Kohn:1952,kohn:1953}, and many others. Here we follow the method of Jost and Kohn. 
A particularly simple model potential results 
in when the phase shift $\delta_0(k)$ is given by effective range expansion  
\bea 
\cot \delta_0(k) = - \frac{1}{k a_s} + \frac{1}{2} r_0 k + \cdots 
\eea 
at low energy under the conditions $k r_0 < 1$ and $r_0 <\! <  |a_s| $. 
Jost and Kohn \cite{PHYSREV:Kohn:1952} showed that,  for negative $a_s$,  the model potential 
takes the form 
\bea 
V_{-}(r) = - \frac{4 \hbar^2}{ \mu r_0^2}  \frac { \alpha \beta^2 \exp(-2 \beta r/r_0) }{ [ \alpha + \exp(-2 \beta r/r_0) ]^2} 
\label{negative}
\eea
where $\alpha = \sqrt{1 - 2 r_0/a_s}$,  $\beta =  1 + \alpha $ and $\mu$ is the reduced mass. 
 This is 
valid for $|\delta_0(E)| < \pi/2$ in the limit $E \rightarrow 0$. When a scattering resonance occurs,  $k |a_s| > \! > 1$ such that 
if one neglects the effects of effective range then $\delta_0 \simeq \pi/2$.  This means that the  
$S$-matrix element $\exp[2 i \delta_0] \simeq - 1$ leading to unitarity-limited 
interactions. We define unitarity regime  by  $-1 <\!< (k a_s)^{-1} <\!< 1 $.

When $a_s$ is positive and large, the potential can support one near-zero energy bound state. Therefore, to obtain a model 
potential for positive $a_s$ from effective range expansion, one needs to incorporate the binding energy of the bound state.
Then the potential becomes a three-parameter potential. This potential is not unique unless 
an additional parameter corresponding to the bound state is taken into account. As shown by Jost and Kohn \cite{kohn:1953},
the $s$-wave binding 
energy $E_b = - \hbar^2 \kappa_0^2/2\mu$ (where $\kappa_0 >0$)  can be parametrized, under effective range expansion, 
by introducing a parameter $\Lambda$ to express $\kappa_0$ in the following form 
\bea 
\kappa_0 = \frac{1}{r_0} \left [ 1 + \sqrt{1 - 2 r_0/a_s} \right ] \frac{1 + \Lambda} {1 - \Lambda} 
\label{k0}
\eea 
It is bounded by $-1 < \Lambda < 1 $ for  $a_s > 2 r_0$  and $r_0 > 0$. Using these three parameters $a_s$, $r_0$ and $\Lambda$, an expression for the 
model potential is given in Eq. (2.29) of Ref. \cite{kohn:1953}, where $\alpha$ is bounded by  
$0 < \alpha \le 1$.  Now, if we make a choice  $\Lambda = -\sqrt{1-2r_0/a_s}$
the Eq.  (2.29) of Ref. \cite{kohn:1953} reduces to a  two-parameter potential of the form 
\bea 
V_{+}(r)=   - \frac{4 \hbar^2}{ \mu r_0^2}  \frac { \alpha \beta^2 \exp(-2 \beta r/r_0) }{ [ 1 + \alpha \exp(-2 \beta r/r_0) ]^2}.  
\label{positive}
\eea
This choice of  $\Lambda$ corresponds to the binding energy $E_{\rm{bin}} \simeq \hbar^2/(2 \mu a_s^2)$ for $2 r_0/a_s <\!<1$. 

It is easy to  notice that  
in the limit $a_s \rightarrow \pm \infty$, both the potential of Eqs. (\ref{negative})  (\ref{positive}) reduces to the form
\bea 
V_{\infty} =   -\frac{4\hbar^2}{\mu r_0^2 \cosh^2(2r/r_0)}
\label{infty}
\eea
This form of the potential has been  employed  by Carlson {\it et al.}  \cite{prl:2003:Carlson} 
for quantum Monte Carlo simulation of a homogeneous  superfluid Fermi gas with infinte  negative scattering length.
Shea et al. \cite{shea:ajp:2009} have calculated the energy spectrum of two harmonically trapped 
 atoms interacting via the potential of Eq. (\ref{infty}), showing that bound state spectrum remains almost the same as for  
 zero-range pseudopotential  \cite{foundphys:1998:butsch} if $r_0$ is much less than the characteristic length scale $l_{ho}=\sqrt{\hbar/\mu \omega_{ho}}$ 
 of the  harmonic oscillator trap with frequency $\omega_{ho}$. 
 
 The potentials of Eqs.  (\ref{negative}) and (\ref{positive})  can  be written in the forms
$ V_{\pm}(r) = V_{0}(r) + V_{\pm}^{(\epsilon)}(r)$ where
\bea 
V_{0}(r) = - \frac{\hbar^2 \kappa^2}{\mu} \frac{\alpha^{-1} }{\cosh^2[\kappa r]}
\label{vptp}
\eea 
and  
\bea 
V_{\mp}^{(\epsilon)} 
&=&  V_{0} \sum_{n=1}^{\infty} (-1)^n (n+1) \left [ \frac{\epsilon}{1 + \exp(\pm 2 \kappa r)} \right ]^n. 
\eea
$V_0(r)$ is in the form of P\"{o}schl-Teller potential of second kind.  
P\"{o}schl-Teller potentials and their different variants are well-known in quantum mechanics 
\cite{book:morsefeshbach,fluegge} as exactly solvable  potentials   in one dimension.  
In three dimension (3D), $s$-wave bound  and continuum (scattering)  solutions of P\"{o}schl-Teller 
potentials are obtained by a group theoretic algebraic approach \cite{barut1,barut2} as well as by non-algebraic methods 
\cite{fluegge,pra:1978:nieoto}. The Schroedinger equation of relative motion can be expressed  in the form 
\bea 
{\mathscr L}_{r} \psi_{\pm}(r) = -  V_{\pm}^{(\epsilon)} \psi_{\pm}(r) 
\label{inhomo} 
\eea
where 
\bea 
{\mathscr L}_r
&=& - \frac{\hbar^2}{2 \mu } \left [ \frac{d^2}{d r^2}  + k^2 \right ]  + V_{0}
\eea
By treating the right hand side of Eq. (\ref{inhomo}) as a source term, we seek  solutions  of 
the homogeneous equation
\bea
{\mathscr L}_{r} \psi_{\pm}(r) = 0
\label{homo}
\eea 
In the following section we present exact solution of $V_0(r)$. 

\section{Solutions of $V_0(r)$}  
 
To obtain solutions of Eq. (\ref{homo}), we first convert this equation into the standard equation for associated Legendre functions. We then write down the 
desired solutions as a superposition of two linearly independent associated Legendre functions by imposing the boundary conditions for 
scattering states. Next, we rewrite the solutions in terms of hypergeometric functions for asymptotic analysis. We then construct Green function 
of Eq. (\ref{homo}). 
In the following subsection we briefly describe our method of solution. In passing it is worth noting that  as the range $r_0$ 
goes to zero, P\"{o}schl-Teller potential in 1D reduces to a delta well potential \cite{jphysa:1999:negro}. 
So, it is expected that for $s$-wave interactions 
three dimensional P\"{o}schl-Teller potential will behave as a contact potential.

\subsection{Scattering solutions of $V_{0}$}

 Let  $\kappa = \beta/r_0$,  $\kappa r = \rho $,  $\cos \theta = \tanh \rho = z $,
then  we have 
\bea 
{\mathscr L}_{\rho} &=& - \frac{\hbar^2}{2 \mu } \left [ \frac{d^2}{d \theta^2} + 
\frac{\cos\theta}{\sin\theta} \frac{d}{d\theta}   - \frac{q^2}{\sin^2\theta} \right ]\sin^2\theta + V_{0}
\eea
 where 
 \bea 
 q=i k/\kappa = \sqrt{- 2 \mu E/\hbar^2}/\kappa. 
 \label{q} 
 \eea 
 Furthermore, substituting $V_{0}$ given by Eq. (\ref{vptp}) in Eq. (\ref{homo}), we have 
\bea 
 \left [ (1 - z^2) \frac{d^2}{d z^2} - 2 z \frac{d}{d z} + 
  \lambda (\lambda + 1)   - \frac{q^2}{1 - z^2} \right ] \psi(z) = 0.
\eea 
where the parameter $\lambda$ is given by  $\lambda (\lambda + 1) = 2 \alpha^{-1} $.
This is  the familiar equation of associated Legendre functions. 
Two linearly independent basic solutions of this equation are   $P_{\lambda}^{q}(z)$ and $Q_{\lambda}^{q}(z)$. 

By doing asymptotic analysis of the two basic solutions as given in the appendix, one can write down the regular scattering solution 
of the potential $V_{0}$ in the form 
\bea 
\psi(z) &=& {\mathscr N}_{\lambda,q} \left [ \Gamma \left ( \frac{1}{2} + \frac{\lambda}{2} 
 +   \frac{q}{2} \right ) P_{\lambda}^{q}(z) 
\right. \nonumber \\ &-& \left.  2 e^{-i \pi q - i \frac{\pi}{2}(q-\lambda -1)} Q_{\lambda}^{q}(z)\frac{1}
{\Gamma \left ( \frac{1}{2} - \frac{\lambda}{2} - \frac{q}{2} \right )} \right ]
\eea 
with ${\mathscr N}_{\lambda,q} = A_{q,\lambda}^{-1}$ where $A_{q,\lambda}$ is given in the appendix.  
One can  notice that
$\alpha$ is bounded by  $ 1 \le \alpha < 2$ or  
 $0 < \alpha \le 1$ depending on whether $a_s$ is  negative or positive, respectively. 
In the limits $a_s \rightarrow \pm \infty$, $\alpha \rightarrow 1$ and therefore $\lambda$ assumes integer 
values of either -2 or 1.

\subsection{Green function }

Although Green functions for one-dimensional 
 P\"{o}schl-Teller potentials are studied before \cite{jmp:1992:kleinert}, to the best of our knowledge the green functions of 
  P\"{o}schl-Teller potentials in  three dimension (3D) is not considered before. 
We first note down the Wronskian \cite{abram-stegun}
\bea 
W\left \{ P_{\lambda}^{q}(z), Q_{\lambda}^{q}(z) \right \}= e^{i \pi q} 2^{2 q} 
\frac
{\Gamma(1+\frac{\lambda + q}{2}) \Gamma(\frac{1}{2} + \frac{\lambda + q}{2})  } 
{(1 - z^2) \Gamma(1+\frac{\lambda - q}{2}) \Gamma(\frac{1}{2} + \frac{ \lambda - q}{2})}
\eea 
between $P_{\lambda}^{q}(z)$ and  $Q_{\lambda}^{q}(z)$. To construct the Green function, we 
prepare another state $\phi(z)$ by linear superposition between  $P_{\lambda}^{q}(z)$ and  $Q_{\lambda}^{q}(z)$ such that 
this state satisfies the outgoing  boundary condition  $\phi(r\rightarrow \infty) \sim - e^{i (k r + \delta_0)}$. 
This leads to 
\bea 
\phi (z) = - \frac{ \pi e^{-i \pi (\lambda + q/2)}}{\Gamma(q) \left [ a_{\lambda}(q) - a_{\lambda}(-q) 
\right ] } e^{i \delta_0} P_{\lambda}^q
\eea 
where 
\bea 
a_{\lambda}(q) = \cos[\pi(\lambda + q)/2]\sin[\pi (q - \lambda)/2] 
\eea 
The Green function can be written in the form  
\bea 
G_{E}(\rho,\rho') \equiv G_{E}(z,z') = - \pi \psi(z_{<})\phi(z_{>})
\eea
That the Green function $G_{E}(\rho,\rho')$ is the solution of the Green equation 
\bea 
{\mathscr L}_{\rho} G_E(\rho,\rho') = - \frac{\hbar^2 (1 - z^2) }{2\mu  } \delta(z - z') =  - \frac{\hbar^2 }{2\mu  } \delta(\rho - \rho')
\eea 
can be ascertained from the relation 
\bea 
\int_{\rho' - 0^+}^{\rho' + 0^+} d \rho {\mathscr L}_{\rho} G_E(\rho,\rho') = 1  
\eea

 \begin{figure}
\includegraphics[width=3.5in]{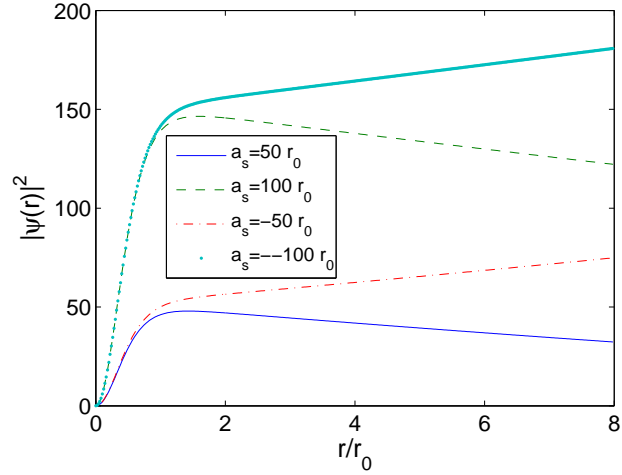}
\caption{Absolute square of energy-normalized continuum wave function $\psi(r)$ in unit of $E_0^{-1} r_0^{-1}$ is plotted 
against $r$ (in unit of $r_0$) for $a_s = 50 r_0$ (solid line), $a_s = 100 r_0$ (dashed lines), $a_s = - 50 r_0$ (dashed-dotted) and $a_s = -100 r_0$ (dotted) 
for  $k = 0.01 r_0^{-1}$. } 
\end{figure}

\subsection{Bound state solutions of $V_{0}$}
 
A bound state occurs for energy $E = - |E|$ or $k = i |k|$, and so the parameter $q$ defined in Eq. (\ref{q}) 
becomes $q = - \bar{k} = - |k|/\kappa$. 
For the solution $\psi_{\pm}^{0}(r)$ to behave as a bound state, asymptotically $\psi (r \rightarrow \infty) \sim e^{-|k| r} $. 
For negative energy, the asymptotic form of 
$\psi_{+}^{0}(r)$  takes the form 
\bea
\psi ( r \rightarrow \infty) \sim  \left [ G_{\lambda}( q) e^{-  |k| r} + G_{\lambda}(-q) e^{|k| r} \right ]
\eea
Now,  for $\psi$ to qualify as a bound state, 
the coefficient of $e^{|k| r}$ must vanish. Thus,  the bound state condition 
is given by 
\bea 
 G_{\lambda}(\bar{k}) = \frac{ 2^{\bar{k}}
\Gamma (  \bar{k}  )}{ \Gamma \left (\frac{1}{2} - \frac{\lambda}{2} + \frac{\bar{k}}{2} \right ) 
\Gamma \left ( 1 + \frac{\lambda}{2} + \frac{\bar{k}}{2} \right )} = 0
 \eea 

Now, assuming   $x = - (\frac{\lambda}{2} + \frac{\bar{k}}{2})$, and using the identity 
$\Gamma(1 - x)\Gamma(x) \pi/\sin[\pi x]$, we can rewrite the above equation in the form 
\bea 
 G_{\lambda}(\bar{k}) = \frac{ 2^{\bar{k}}}{\pi} \frac{ \Gamma \left ( -  \frac{\lambda}{2} - \frac{\bar{k}}{2} \right ) }
 { \Gamma \left (\frac{1}{2} - \frac{\lambda}{2} + \frac{\bar{k}}{2} \right )} 
 \sin  \pi \left ( -  \frac{\lambda}{2} - \frac{\bar{k}}{2} \right ) 
 \label{glamda}
 \eea
This shows that  $G_{\lambda}(\bar{k}) = 0$ when 
\bea 
- \lambda - \bar{k} = 2 n 
\label{bcond}
\eea 
with  $n$ being an integer, or the argument of the gamma function in the denominator on the right hand side of the Eq. (\ref{glamda}) 
is a negative integer. As $a_s$ decreases  from $+\infty$ to $2r_0$, $\lambda$ decreases 
from -2 to $-\infty$. This means that the argument of this gamma function can not be negative since $\bar{k} \ge 0$.  
Therefore, the bound state condition is given by Eq. (\ref{bcond}) which, 
  for low energy,  will be satisfied for $n=1$. On the other hand, when $a_s < 0$ and as 
$a_s$ changes from $-\infty$ to small  negative value  $\lambda$ decreases from 1 ranging between 
$0 < \lambda \le 1$. This means 
the above equation can not be fulfilled for negative scattering length. 
These findings are consistent with the conditions for bound states obtained earlier by group theoretic approach 
\cite{barut2}. 

\begin{figure}
\includegraphics[width=3.5in]{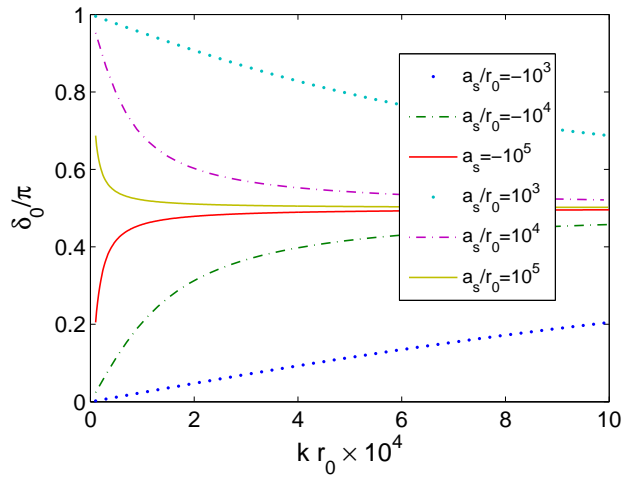}
\caption{The phase shift $\delta_0(k)$ in unit of $\pi$ radian is plotted against dimensionless momentum $k r_0$ (scaled by a factor 
of $10^4$) for three positive and three negative values of  scattering length $a_s$ as indicated in the legend. The lower dotted, dashed and solid 
curves correspond to $a_s = - 10^3$, $a_s = - 10^4$ and $a_s = - 10^5$, respectively; while the upper  dotted, dashed and solid 
curves  are plotted for positive $a_s$ of same magnitudes, respectively.  } 
\end{figure}

\section{Results and discussions} 

In figure 1, we show the square of energy-normalized scattering or continuum wave function $|\psi(r)|^2$ as a 
function of $r$ for four values of relatively large scattering lengths at a fixed low energy. The contrasting 
behavior of the wave functions for positive and negative scattering lengths for $r \ge r_0$ is noteworthy. The 
intercepts of the slope of the wave function at $r = r_0$ on the $x$-axis are positive and negative for positive 
and negative scattering lengths, respectively. The intercepts are nearly equal to the respective scattering lengths. 
Thus the wave functions behave exactly like those of a 3D square well at  low energy. At $r=r_0$, the wave functions for
positive and negative  scattering lengths bend towards and away from $x$-axis, respectively.  

Next we analyze  phase shifts at low energy. From the asymptotic analysis of 
the solution for $V_0$ as given in the appendix,   we have 
$\psi(r \rightarrow \infty)  \sim \sin[k r + \delta_0]$ where  
\bea 
\delta_0(k)  &=&   - \tan^{-1} \left [ \cot \frac{-\pi \lambda }{2} \tanh \left ( \frac{\pi k}{2} \right ) \right ]
+   \arctan \left ( - \frac{k}{\lambda \kappa} \right )   \nonumber \\ 
&+&  \sum_{n=1}^{\infty} \left [ 
 \arctan \left (\frac{k }{(-\lambda + n) \kappa } \right )  
 -   \arctan \left ( \frac{k }{ n \kappa} \right )\right ] 
 \label{phaseshift}
\eea 
From the expression for $\delta_0$ given in Eq. (\ref{phaseshift}) one can notice that, for $k \ne 0$, 
as $\lambda \rightarrow 1$ we get $\delta_0 \rightarrow \pi/2 - 0_+$ while as $\lambda$ approaches to -2, 
$\delta_0$ approaches to $\pi/2 + 0_+$. 
From the relation 
 $a_s = - \lim_{k \rightarrow 0} \tan [\delta_0(k)]/k $, one finds  that, for  negative $a_s$ 
the resonance ($a_s \rightarrow - \infty$) will be characterized 
by $\delta_0 \rightarrow \pi/2 - 0_+$, that is, by $\lambda = 1$ while for positive $a_s$ the resonance will occur 
when $\lambda = -2$. The Eq. (\ref{phaseshift}) suggests that in order for $\delta_0(k) \le \pi/2$ so that $a_s < 0$, one needs 
 to set  $\lambda = \lambda_{1} = - 1/2 +  \sqrt{ 1 + 8 \alpha^{-1} }/2$. Similarly, for $a_s > 0$ one has to fix   
 $\lambda = \lambda_{2} = - 1/2 -  \sqrt{ 1 + 8 \alpha^{-1} }/2$. This analysis shows that  we have $0 < \delta_0 < \pi/2$ and 
  $\pi/2 < \delta_0 < \pi$ for $a_s <0$ and $a_s > 0$, respectively. 
 For $|\lambda| <\!< 1$ the phase 
shift $\delta_0 (k) \propto k $ in accordance with Wigner's threshold laws.  
In the limit $\alpha \rightarrow 1 $, $\lambda$ can assume a value of either -2 or 1. In both limits 
$a_s \rightarrow - \infty$ 
and $a_s \rightarrow + \infty$, the parameter $\lambda$ is given by 
$\lambda (\lambda + 1) = 2 $. 
 
Figure 2 displays the variation of $\delta_0(k)$ as a function of $k r_0$ for different values of $a_s$. To remain within 
 the validity regime of the effective range expansion, in this plot we have restricted the variation of $k r_0$ at low values 
 below $10^{-3}$. From this figure we notice that for large scattering length $\delta_0(k)$ approaches the resonant value 
 $\pi/2$ at finite momentum while in the zero momentum limit $\delta_0(k)$ significantly deviates from $\pi/2$ even for large scattering 
 length. This implies that  the momentum- and range-dependence of interactions become particularly important for  resonant interactions.

We next discuss how these potentials will be useful to model tunable MFR in ultracold atoms.  
The $s$-wave phase shift near MFR is $\delta_0 (k) = \delta_{{\rm bg}} + \delta_r$ where 
the background phase shift $ \delta_{{\rm bg}}$ can be approximated as  $\delta_{{\rm bg}} \simeq -k a_{{\rm bg}}$ 
with $a_{{\rm bg}}$ being the background scattering length.  
Here  $\delta_r$ is the resonance phase shift. At low energy, the Feshbach resonance width $\Gamma_f$ is given by 
$\Gamma_f/2 \simeq k a_{bg} \Gamma_0$ where $\Gamma_0$ is a parameter 
related to the width of zero crossing.  
 Now, assuming $|k a_{bg}| <\!< 1$ one can write  
 \bea 
\frac{1}{a_s} =   \frac{B_0 - B}{\Delta a_{bg}} 
\eea 
and 
\bea 
r_0 = 2 a_{{\rm bg}} - \frac{\hbar}{ \mu a_{{\rm bg}} \Gamma_0}
\label{r0}
\eea 
where $B_0$ is the magnetic field at which MFR occurs (or equivalently, $a_s$ diverges)
at zero energy.
Since $a_{{\rm bg}} \Gamma_0 >0$, the above equation shows that $r_0$ will be positive if  $ a_{{\rm bg}}$ is positive and $ a_{{\rm bg}} > \sqrt{ \hbar/ (2 \mu \Gamma_0) }$ . If $a_{bg} < 0$ then 
$r_0$ is negative. Since the model potentials of Eqs. (\ref{negative}) and (\ref{positive}) are derived assuming $r_0 > 0$, 
these potentials will be useful to model those MFR for which $r_0$ is positive. For example,
magnetic Feshbach Resonances observed in $^{133}$Cs near $B_0 = -11.7, 547 $ 
and 800 G  fulfill the conditions 
for $r_0 > 0 $ \cite{chin:2004:pap1,chin:2004:pap2}. In general, 
for a relatively broad Feshbach resonance or an open-channel dominated Feshbach resonance \cite{rmp:2010:chin} with positive $a_{{\rm bg}}$ 
for which  binding energy of closed channel bound state has the universal form $E_{b} \sim \hbar^2/(2\mu a_s^2)$, the resonance can 
be described effectively  by a single-channel potential using our proposed model potentials. The treatment of MFR with our proposed finite-range 
potentials will give new insight into resonant phenomena in ultracold atoms by providing effective range dependence of scattering phase shifts 
and near-zero energy bound states. The analytical solutions found in this paper will be useful to calculate photoassociation in the presence of 
an MFR and thereby to develop quantitative understanding of the Fano effect in photoassociation \cite{gsa:jpb:2009}. Furthermore, in case of 
two-component Fermi gas of atoms, the modeling of MFR with the finite-range potentials will facilitate to investigate  the hither-to-unexplored 
nontrivial effects of large effective range on the BCS-BEC crossover from Bardeen-Cooper-Schrieffer (BCS) state of Cooper-pairs to Bose-Einstein 
condensate (BEC) of dimers.    

\section{Conclusions and outlook} 

In conclusion we have demonstrated that the effective range dependence of resonant two-body interactions can be taken into account 
within a class of  model potentials constructed under effective range expansion using the methods of Jost and Kohn. 
We have shown that these potentials reduce to the form 
of P\"{o}schl-Teller potential when scattering length is large. We have presented analytical scattering and bound state solutions 
of the potentials for large scattering lengths and established the connections of the nature of the solutions with the sign and strength 
of the $\lambda$-parameter of the P\"{o}schl-Teller potential. These finite-range model potentials will permit us to explore the 
finite-range effects of interactions between atoms in low-dimensional traps. In a recent paper \cite{ijp:2015:partha}, using these 
potentials we have numerically studied  bound-state properties of two atoms in a quasi-two dimensional trap and found significant effects 
of the range  on the bound states. For studies of collisional properties of cold atoms near resonances, one can 
employ more accurate multichannel scattering-based computational methods where one can use molecular potentials. However, 
to gain insight into  many-body physics near resonances, one prefers  a single-channel simplified model potential that can be Fourier 
transformed so that one can conveniently develop many-body treatment of a homogeneous system in momentum space. Keeping 
this in mind, we have shown  that there exists a class of simplified finite-range potentials that are  well suited for the purpose of describing 
resonant interactions more accurately with both finite-range and energy-dependence of the scattering processes being taken into account.

\appendix
\section{Asymptotic analysis}  

The regular scattering solution is required to fulfill the boundary conditions $\psi(r=0) = 0$ and 
$\psi(r \rightarrow \infty) \sim \sin[ k r + \delta_0(k)]$ or equivalently, $\psi(z=0) = 0$ and 
$\psi(z \rightarrow 1) \sim \sin[ \kappa \rho + \delta_0(k)]$. This means  $\psi(z)$ will be given 
by the superposition of the two  basic solutions fulfilling these two boundary conditions. Thus we have 
\bea 
\psi(z) \sim \Gamma \left ( \frac{1}{2} + \frac{\lambda}{2} + \frac{q}{2} \right ) P_{\lambda}^{q}(z) 
- 2 e^{-i \pi q - i \frac{\pi}{2}(q-\lambda -1)} Q_{\lambda}^{q}(z)\frac{1}
{\Gamma \left ( \frac{1}{2} - \frac{\lambda}{2} - \frac{q}{2} \right )} \nonumber \\
\label{regular} 
\eea 
Expressing  the two associated Legendre functions in terms of hypergeometric functions, and after a lengthy algebra  
one obtains  
\bea
\psi(z) &\sim&  2 i 
2^q e^{ i \pi(\lambda+q)} \pi^{1/2}  
\frac{\Gamma \left ( \frac{1}{2} + \frac{\lambda}{2} + \frac{q}{2} \right )}
{\Gamma \left ( \frac{1}{2} + \frac{\lambda}{2} - \frac{q}{2} \right )
 \Gamma\left (- \frac{\lambda}{2} - \frac{q}{2} \right ) \sin \left [ \pi \left ( \frac{\lambda}{2} + \frac{q}{2} \right ) \right ]} \nonumber \\
&\times&  z (1 - z^2)^{-1/2}  (z^2 - 1)^{\frac{\lambda}{2}} F\left ( \frac{1}{2} - \frac{\lambda}{2} - \frac{q}{2}, \frac{1}{2}  -
\frac{\lambda}{2} + 
\frac{q}{2}; \frac{3}{2}; \frac{z^2}{z^2 - 1}   \right ) 
\eea 
From this expression it is easy to verify that at $r=0$ or $z=0$, $\psi = 0$ and therefore  $\psi(z)$ 
is a regular solution. 

Now, we will show that $\psi(z)$ satisfies the proper asymptotic boundary condition 
$\psi (r \rightarrow \infty) \sim \sin[k r + \delta_0(k)]$.  
Since $z^2 = \tanh^2 \rho$,  $\frac{z^2}{z^2 - 1} = -\sinh^2 \rho$, we can write  
$z (1 - z^2)^{-1/2}  (z^2 - 1)^{\frac{\lambda}{2}} =  e^{i \pi \lambda/2} \sinh \rho \left [ \cosh^2\rho \right ]^{-\lambda/2}$ 
and 
\bea 
&&F \left ( \frac{1}{2} - \frac{\lambda}{2} - \frac{q}{2}, \frac{1}{2} - \frac{\lambda}{2} + 
\frac{q}{2}; \frac{3}{2}; 
 \frac{z^2}{z^2 - 1}  \right ) =
\frac{\Gamma\left (\frac{3}{2}\right )
\Gamma (  q  )}{ \Gamma \left (\frac{1}{2} - \frac{\lambda}{2} + \frac{q}{2} \right ) 
\Gamma \left ( 1 + \frac{\lambda}{2} + \frac{q}{2} \right )} \nonumber \\
&\times&
( \sinh^2 \rho)^{ \frac{\lambda}{2} + \frac{q}{2} - \frac{1}{2}  } 
F\left ( \frac{1}{2} - \frac{\lambda}{2} - \frac{q}{2}, 
- \frac{\lambda}{2} - \frac{q}{2}; 1 - 
q; - \frac{1}{\sinh^2 \rho} \right )
\nonumber \\
&+&  
\frac{\Gamma\left (\frac{3}{2}\right )
\Gamma ( - q  )}{ \Gamma \left (\frac{1}{2} - \frac{\lambda}{2} - \frac{q}{2} \right ) 
\Gamma \left ( 1 + \frac{\lambda}{2} - \frac{q}{2} \right )} \nonumber \\
&\times& 
( \sinh^2 \rho)^{ \frac{\lambda}{2} - \frac{q}{2} - \frac{1}{2}  }
F\left (\frac{1}{2}  - \frac{\lambda}{2} + \frac{q}{2}, 
- \frac{\lambda}{2} + \frac{q}{2}; 1 + 
q; - \frac{1}{\sinh^2 \rho} \right ) \nonumber \\
\eea 
Thus we have  
\bea
\psi( r \rightarrow \infty) \sim  A_{q,\lambda} \left [ G_{\lambda}(q) e^{i k r} + G_{\lambda}(-q) e^{-i k r} \right ]
\eea
where 
\bea 
A_{q,\lambda} =   i   
\frac{2^{1+q} e^{ i \pi q}  e^{ i \pi 3 \lambda/2}  \pi^{1/2} \Gamma(3/2) \Gamma \left ( \frac{1}{2} + \frac{\lambda}{2} + \frac{q}{2} \right )}
{  \sin \left [ \pi \left ( \frac{\lambda}{2} + \frac{q}{2} \right )  \Gamma \left ( \frac{1}{2} + \frac{\lambda}{2} - \frac{q}{2} \right )
 \Gamma\left (- \frac{\lambda}{2} - \frac{q}{2} \right ) \right ]} 
\eea
and 
\bea 
G_{\lambda}(q) = \frac{ 2^{-q}
\Gamma (  q  )}{ \Gamma \left (\frac{1}{2} - \frac{\lambda}{2} + \frac{q}{2} \right ) 
\Gamma \left ( 1 + \frac{\lambda}{2} + \frac{q}{2} \right )}
\eea 
The quantity within the third bracket on the right hand side of Eq. (A4) can be expressed in the form 
$   |G_{\lambda}(q)| \sin\left [ k r + \delta_0 \right ] $ with $\delta_0 =  \frac{\pi}{2} +  \phi$ where 
\bea 
\phi = {\rm{arg}}\left [ \frac{2^{-q}
\Gamma (  q  )}{ \Gamma \left (\frac{1}{2} - \frac{\lambda}{2} + \frac{q}{2} \right ) 
\Gamma \left ( 1 + \frac{\lambda}{2} + \frac{q}{2} \right )} \right ]
\eea 
Substituting $q = i k/\kappa$  and making use of the standard phase relationship involving gamma functions we obtain  
\bea 
\phi 
&=& {\rm{arg}} \Gamma (i k /\kappa )  - {\rm{arg}} \Gamma(-\lambda + i k/\kappa)
 - \tan^{-1} \left [ \cot \frac{-\pi \lambda }{2} \tanh \left ( \frac{\pi k}{2} \right ) \right ] 
\eea 
Using the formulas [6.1.27], [6.3.16], [6.3.7] of Ref. \cite{abram-stegun}, one can obtain Eq.(26).

\end{document}